\documentclass[a4paper,oneside,english]{aa}
\usepackage{times}
\usepackage[T1]{fontenc}
\usepackage{amsmath}
\usepackage{graphicx}
\usepackage{amssymb}
\usepackage{natbib}
\usepackage{babel}
\usepackage{xspace}
\bibpunct{(}{)}{;}{a}{}{,}
\newcommand{\ascc}   {\mbox{ASCC-2.5}\xspace}

\begin{document}

\title{Why simple stellar population models do not reproduce the colours of Galactic open clusters}

\author{A.E.~Piskunov \inst{1,2,3} \and
        N.V.~Kharchenko \inst{1,3,4} \and
        E.~Schilbach \inst{3} \and
        S.~R\"{o}ser \inst{3} \and
        R.-D.~Scholz \inst{1} \and
        H.~Zinnecker \inst{1} }

\offprints{R.-D.~Scholz}

\institute{Astrophysikalisches Institut Potsdam, An der Sternwarte 16,
D--14482 Potsdam, Germany\\
email: rdscholz@aip.de
\and
Institute of Astronomy of the Russian Acad. Sci., 48 Pyatnitskaya
Str., 109017 Moscow, Russia
\and
Astronomisches Rechen-Institut, M\"{o}nchhofstra\ss{}e 12-14, Zentrum f\"ur Astronomie der
    Universit\"at Heidelberg,
D--69120 Heidelberg, Germany
\and
Main Astronomical Observatory, 27 Academica Zabolotnogo Str., 03680
Kiev, Ukraine
}

\date{Received 7 September 2009 / Accepted 3 October 2009}

\abstract{For Galactic open clusters, fundamental parameters such as age
or reddening are usually determined independently of their integrated colours.
For extragalactic clusters, on the other hand, they are derived by
comparing their integrated colours with predictions of simple stellar
population (SSP) models.}
{We search for an explanation of the disagreement between the observed integrated
colours of 650 local Galactic clusters and the theoretical colours of present-day 
SSP models.}
{We check the hypothesis that the systematic offsets 
between observed and theoretical 
colours, which are $(B$$-$$V)\approx 0.3$
and $(J$$-$$K_s)\approx 0.8$, 
are caused by neglecting the discrete nature of the underlying
mass function. Using Monte Carlo simulations, we 
construct artificial clusters of 
coeval stars taken from a mass distribution 
defined by an Salpeter initial mass function (IMF) and compare them with
corresponding ``continuous-IMF'' SSP models.}
{If the discreteness of the IMF is taken into account, the model fits the
observations perfectly and is able to explain naturally a number of red
``outliers'' observed in the empirical colour-age relation. We find that
the \textit{systematic} offset between the continuous- and
discrete-IMF colours reaches its maximum of about 0.5 in $(B$$-$$V)$ for
a cluster mass $M_c=10^2\,m_\odot$ at ages
$\log t\approx 7$, and diminishes substantially but not
completely to about one hundredth of a magnitude at $\log t >7.9$ at cluster masses
$M_c> 10^5\,m_\odot$. At younger ages, it is still present even in massive
clusters, and for $M_c \leqslant 10^4\,m_\odot$ it is larger than 0.1 mag 
in $(B$$-$$V)$. 
Only
for very massive clusters ($M_c>10^6\,m_\odot$) with ages $\log t< 7.5$ is
the offset small
(of the order of 0.04 mag) and smaller than the typical observational error 
of colours of extragalactic clusters.}
{}


\maketitle

\section{Introduction}
\label{sec:intro}

Using data on accurate and homogeneous spatio-kinematic-photometric membership
for 650 Galactic open clusters, we previously computed their integrated
magnitudes in $B,V,J,H$, and $K_s$-passbands \citep{intpar}. The magnitudes are based
on accurate and uniform data from the \ascc catalogue and were computed by adding
the individual luminosities of the most secure cluster members. In contrast to
previous lists of integrated magnitudes \citep[e.g., those of][]{battin94,
lata02} of Galactic star clusters, our data provide an independent 
and more importantly uniform dataset.
Hence, our integrated magnitudes can and should be used as benchmarks
for studies of the populations of extragalactic star clusters.

\begin{figure}[t]
\resizebox{\hsize}{36mm}{
\includegraphics[clip]{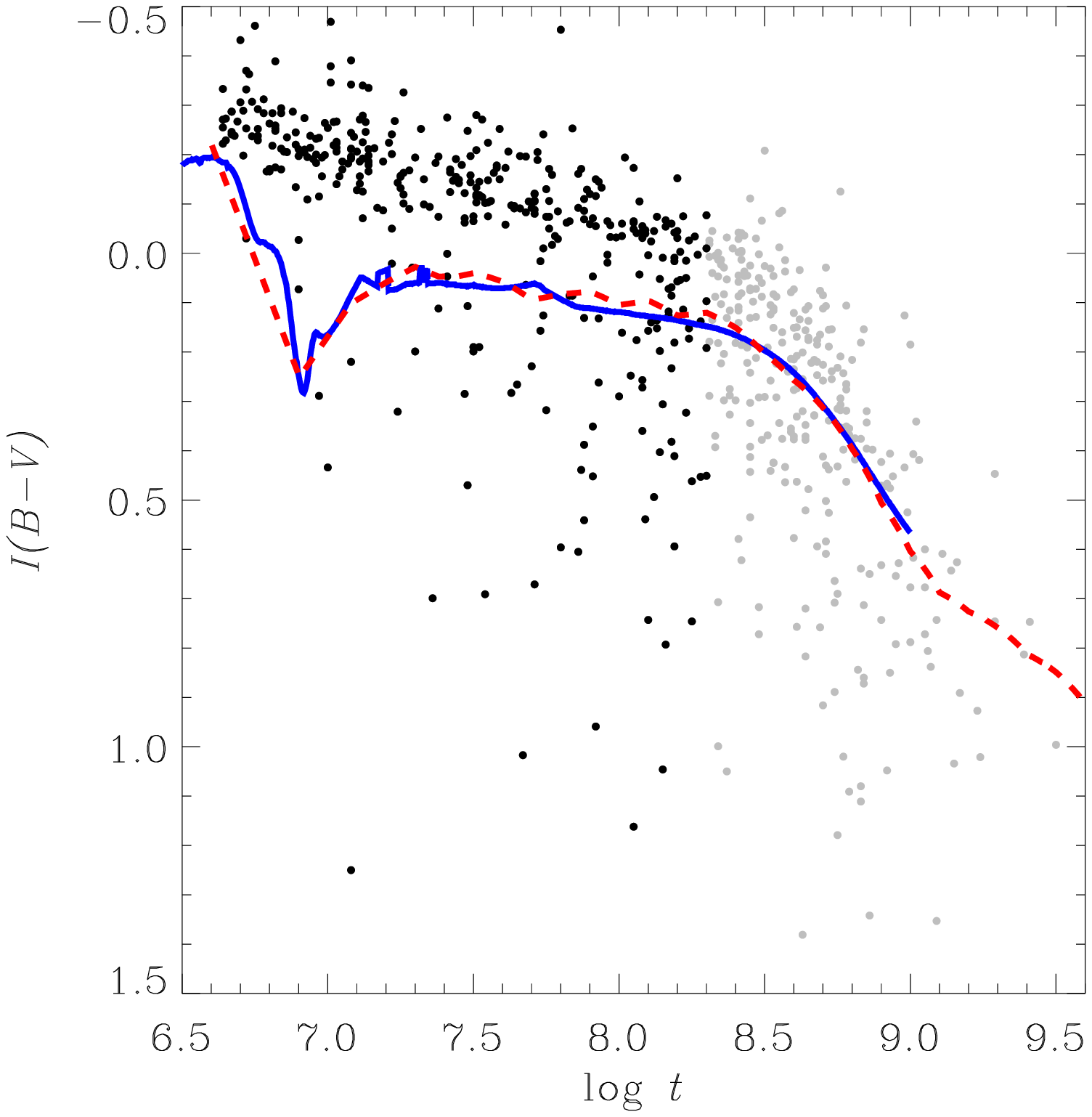}
\includegraphics[clip]{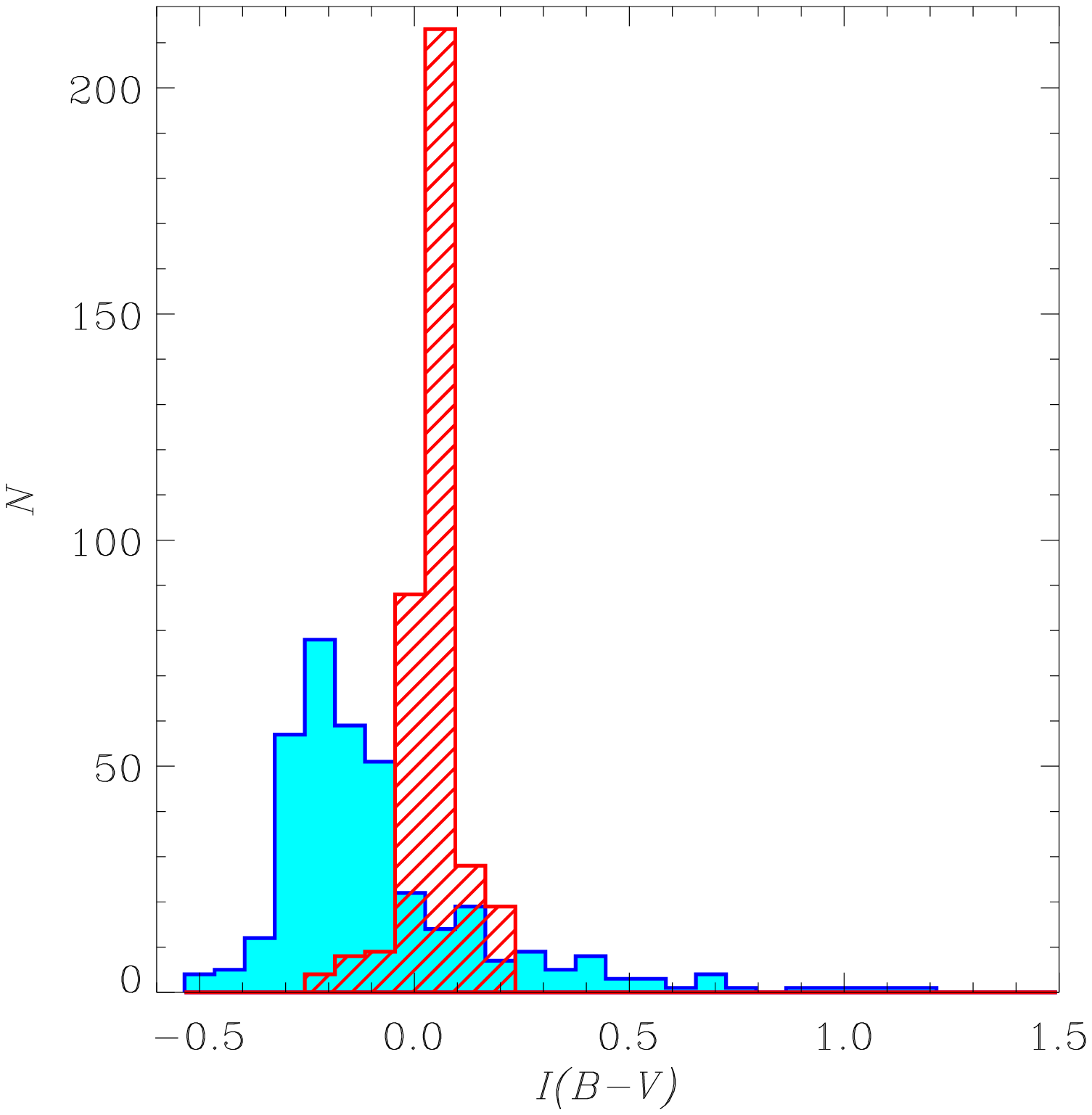}
}\\
\resizebox{\hsize}{36mm}{
\includegraphics[clip]{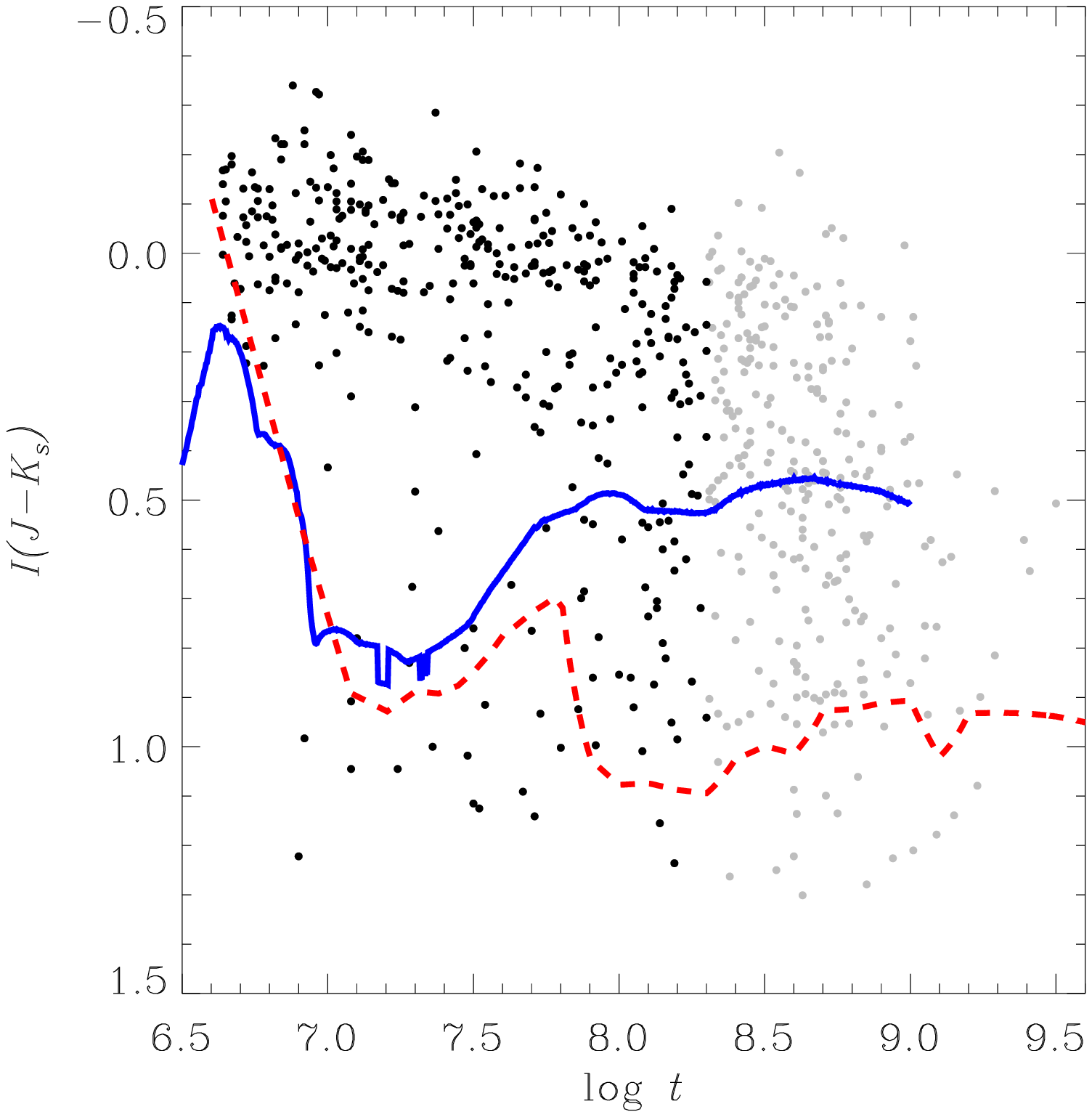}
\includegraphics[clip]{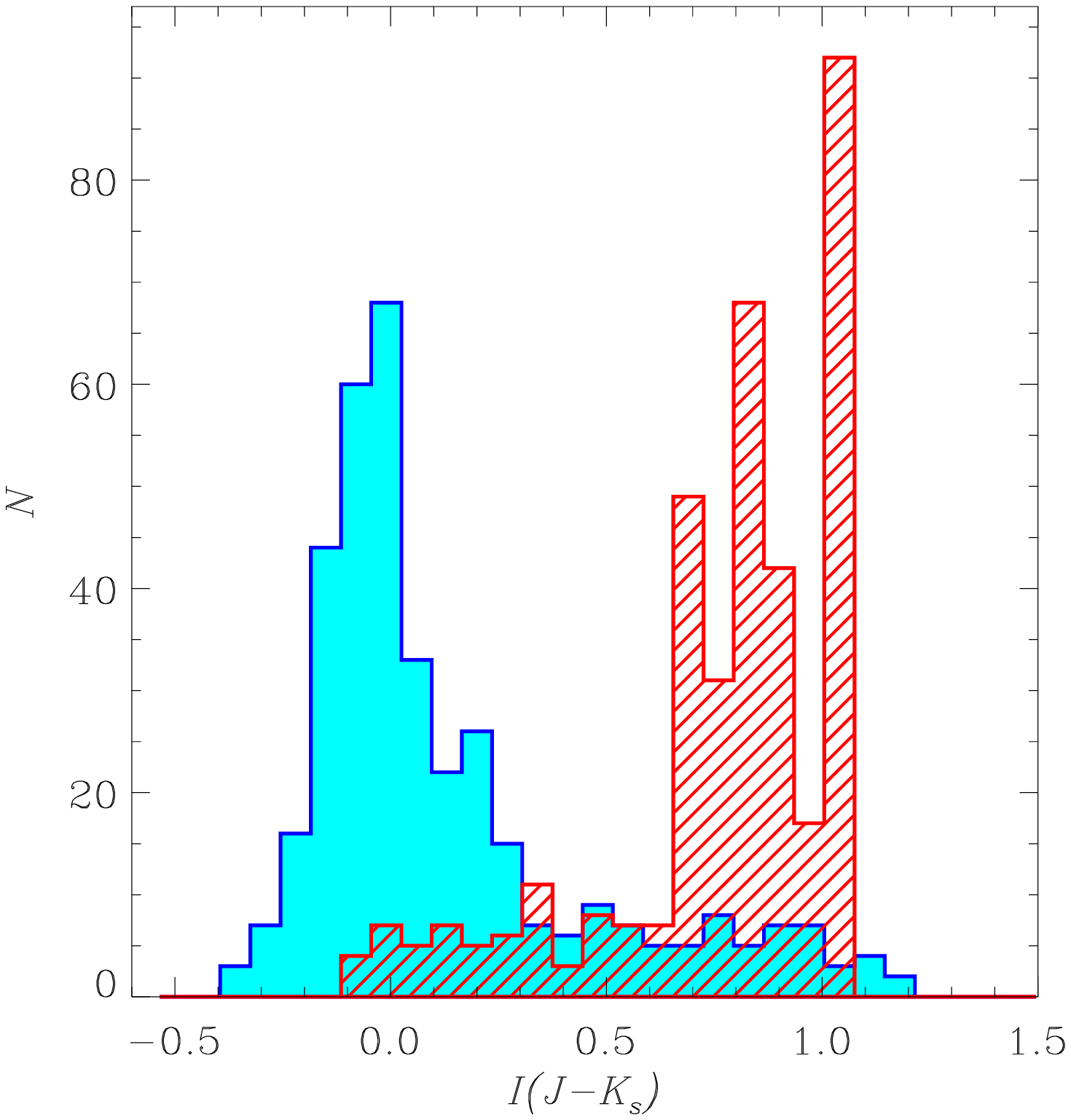}
}
\caption{Observed colours of 
650 Galactic open clusters compared to
theoretical colours computed from standard SSP models. The upper row is 
for $B$$-$$V$, the bottom one for $J$$-$$K_s$. 
The left column shows ``integrated 
colour $vs. \log t\,$'' diagrams, the right one the distributions of colour 
indices. Dots show Galactic open clusters from our sample. Black dots are 
young clusters used for constructing the colour distribution, grey dots the
older clusters. The curves are the SSP model tracks (the solid one is from
Starburst99, the dashed one is from GALEV).The observed colour distributions 
are shown as the
filled cyan histogram, the GALEV model distribution as the hatched red one.
}
\label{sev_fig}
\end{figure}

Based on these uniform data, we found disagreement between the 
observed colours of our cluster sample and theoretical colours derived 
from simple stellar population (SSP) models. The comparison of 
the \citet{intpar} data with present-day SSP
models such as GALEV \citep{galev03} and Starburst99  \citep{sb99_05}  
illustrated a substantial
offset of the order of $0.2-0.3$ mag in $(B$$-$$V)$ of 
the model colours with respect to the observations for clusters
younger than $\log t \lesssim 8.5$. This is both remarkable
and important, since most
data such as age and mass collected nowadays for
the  bulk of extragalactic clusters, are derived by comparing their 
integrated light with the predictions of 
SSP models \citep[see e.g.,][]{bikea03}.
The models are produced by evolutionary synthesis codes simulating the extensive
and complicated stellar populations of galaxies. However, they are also believed 
to be appropriate for describing cluster populations more than five orders of magnitude 
less massive (and apparently far simpler). 
Although star clusters may be influenced
by effects that are irrelevant to galaxies (e.g.,
low number statistics, because of the discreteness of the real IMF, or
dynamical mass-loss by evaporation of stars), the validity of the SSP
approach for star clusters 
can be tested from general considerations \citep{cervino04}. 
In contrast, in this paper we apply the issue
of the discreteness of the IMF to the specific
case of Galactic open clusters.

A colour offset of this kind was found previously by
\citet{lata02}, in the colour indices $U$$-$$B$, $B$$-$$V$, $V$$-$$R$, and $V$$-$$I$.
However, they accept that only the offsets in $V$$-$$R$ and $V$$-$$I$ represent 
significant
deviation of their observations from the model predictions. We suspect that 
the discreteness of the IMF in open clusters may explain these discrepancies.
To study this issue in a more systematic way,
we constructed our own SSP model that takes into account the effect of
low number statistics (discreteness), which is a common problem for open star clusters.
We cross-checked a continuous version of our code with GALEV and found good
agreement.

The discreteness of the IMF itself is a natural assumption for any stellar ensemble 
consisting of
individual objects. However, when one considers vast stellar populations (e.g.,
galaxies) that densely populate the entire colour-magnitude diagram, one can
adopt a continuous IMF, which is more convenient for
different technical reasons. For small populations (stellar clusters), the
assumption of a discrete IMF is clearly appropriate..
This letter
describes the solution of a restricted problem, namely the influence of the discrete-IMF
approach on the colours of young clusters.
The full scope of results related to the
IMF-discreteness effect on star clusters will be reported elsewhere.

\section{The SSP models and observations}
\label{sec:stand}

For the determination of integrated magnitudes (and colours) of Galactic
open clusters, we refer the reader to \citet{intpar}, which also
gives an overview of the cluster sample and references for further
reading.

For simplicity, we refer to our SSP model hereafter as the ``realistic'' model, and
call the literature models ``standard'' models.
The standard SSP models were computed with help of the online servers
provided by the GALEV\footnote{\texttt{http://www.galev.org/}}, and
Starburst99\footnote{\texttt{http://www.stsci.edu/science/starburst99/}} sites.

The input parameters of the realistic model are the total cluster mass $M_c$, the initial
mass function (IMF) of cluster stars
\[
f(m)=dN/dm\,,
\]
($N$ being the number of stars with masses between $m,m+d\,m$), and the lower
and upper limits ($m_l$ and $m_u$)
of the range of the allowed masses of cluster stars. We adopt the Salpeter IMF,
$f(m)=k\,m^{-\alpha},\, \alpha=2.35$, where $k$ 
is a normalisation factor that depends on the assumed cluster mass. The upper mass
limit is confined to the high-mass end of the adopted grid of evolutionary
tracks/isochrones. Only at the very early stages
of cluster evolution does $m_u$ have an impact on integrated colours.
Given a fixed IMF slope, the lower limit governs the
population of the massive end of the IMF at given $M_c$. We consider $M_c$
and $m_l$ here as free parameters of the adopted model. The model stellar sample was
generated from the IMF with a random number generator.

The photometric properties of the model are defined by the implemented grid of the
isochrones. For our model, we have used the grid provided by the Padova group via
the online server CMD\footnote{\texttt{http://stev.oapd.inaf.it/cgi-bin/cmd}}.
According to our
experience, the Padova isochrones fit the brighter parts of the open
cluster CMDs reasonably well over a wide
range of cluster ages. These parts are responsible for the integrated colours.
We used a solar metallicity grid ($Z=0.019$), and retrieved the
passbands $U,B,V,R,I$ and $J,H,K_s$ with an age range $\log t=6.0-10.2$ and a step
in $\log t$ of 0.05. For compatibility, we selected 
the same set of parameters (i.e., underlying IMF, mass range limits,
isochrone set) for the standard
models. However, since the Padova group regularly updates their
isochrones, 
full compatibility between the isochrone grids used by the realistic and
standard models cannot be guaranteed.

In Fig.~\ref{sev_fig}, we compare 
the location of the observations with the curves of the standard model in the
colour-age diagram. GALEV and
Starburst99 tracks agree reasonably well with each other. The above-mentioned 
offset of
the models with respect to observed colours for $\log t\lesssim8.5$
is obvious in both of the colour-age diagrams.

\begin{figure}[t]
\resizebox{\hsize}{36mm}{
\includegraphics[clip]{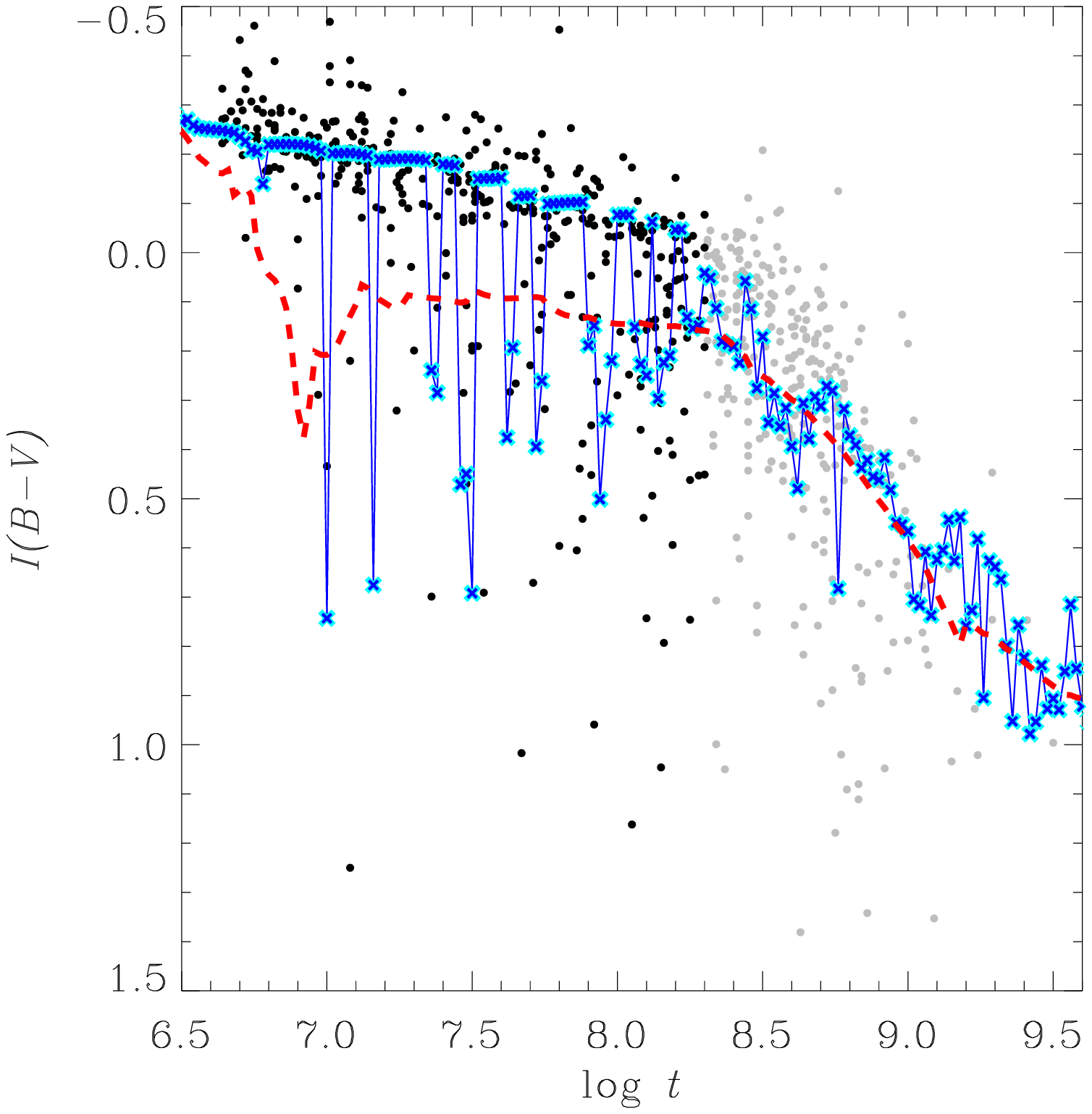}
\includegraphics[clip]{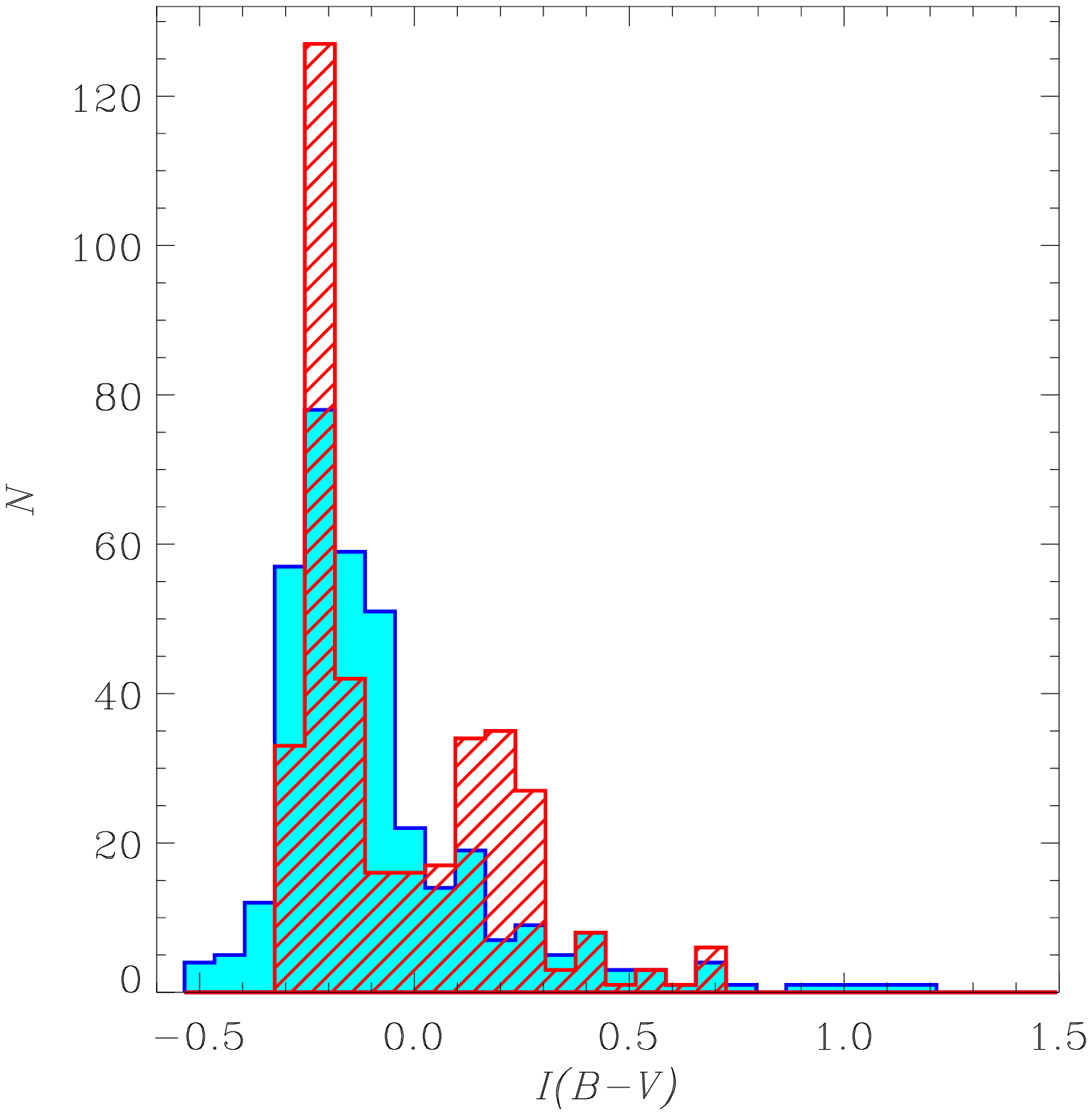}
}\\
\resizebox{\hsize}{36mm}{
\includegraphics[clip]{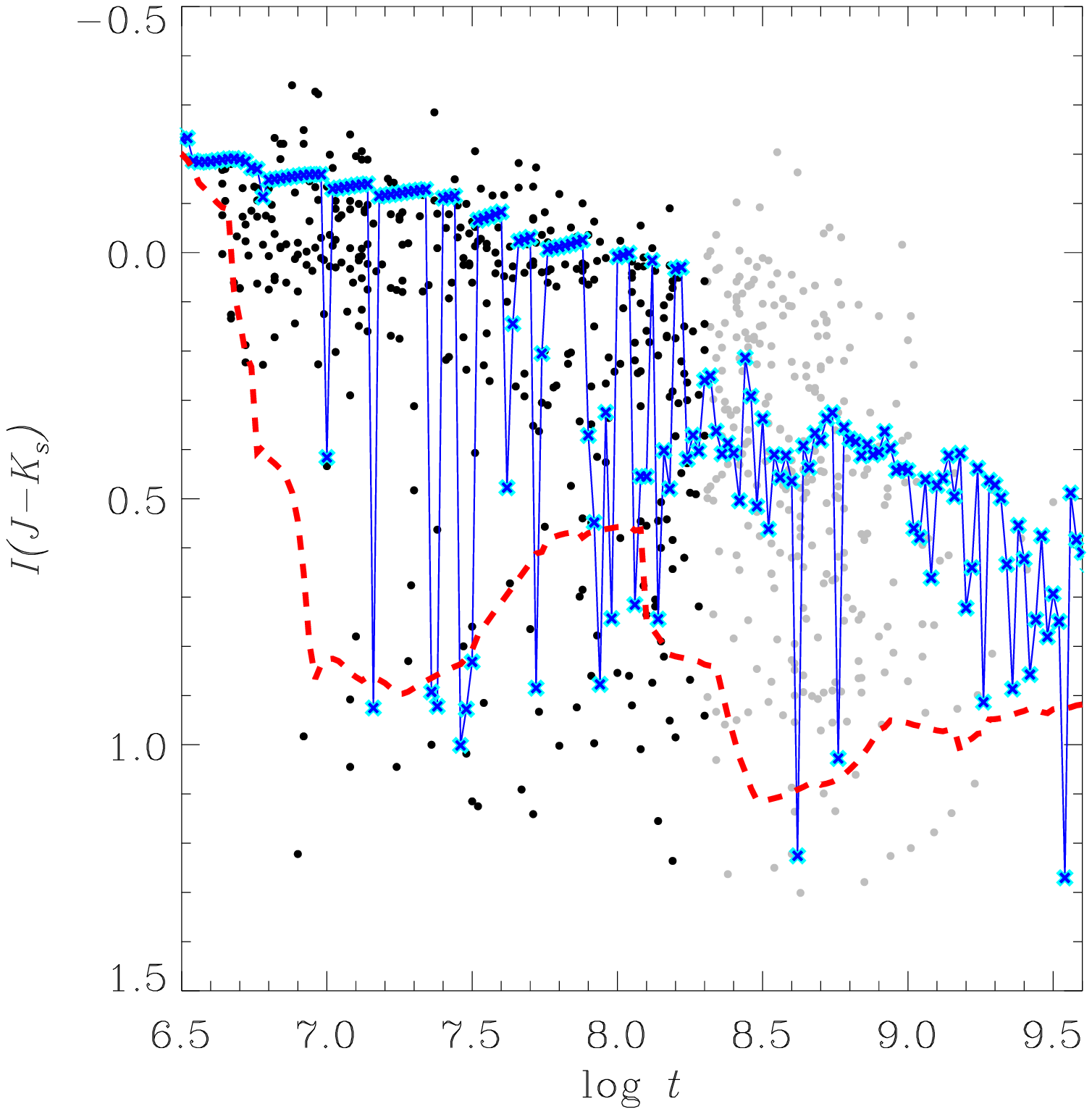}
\includegraphics[clip]{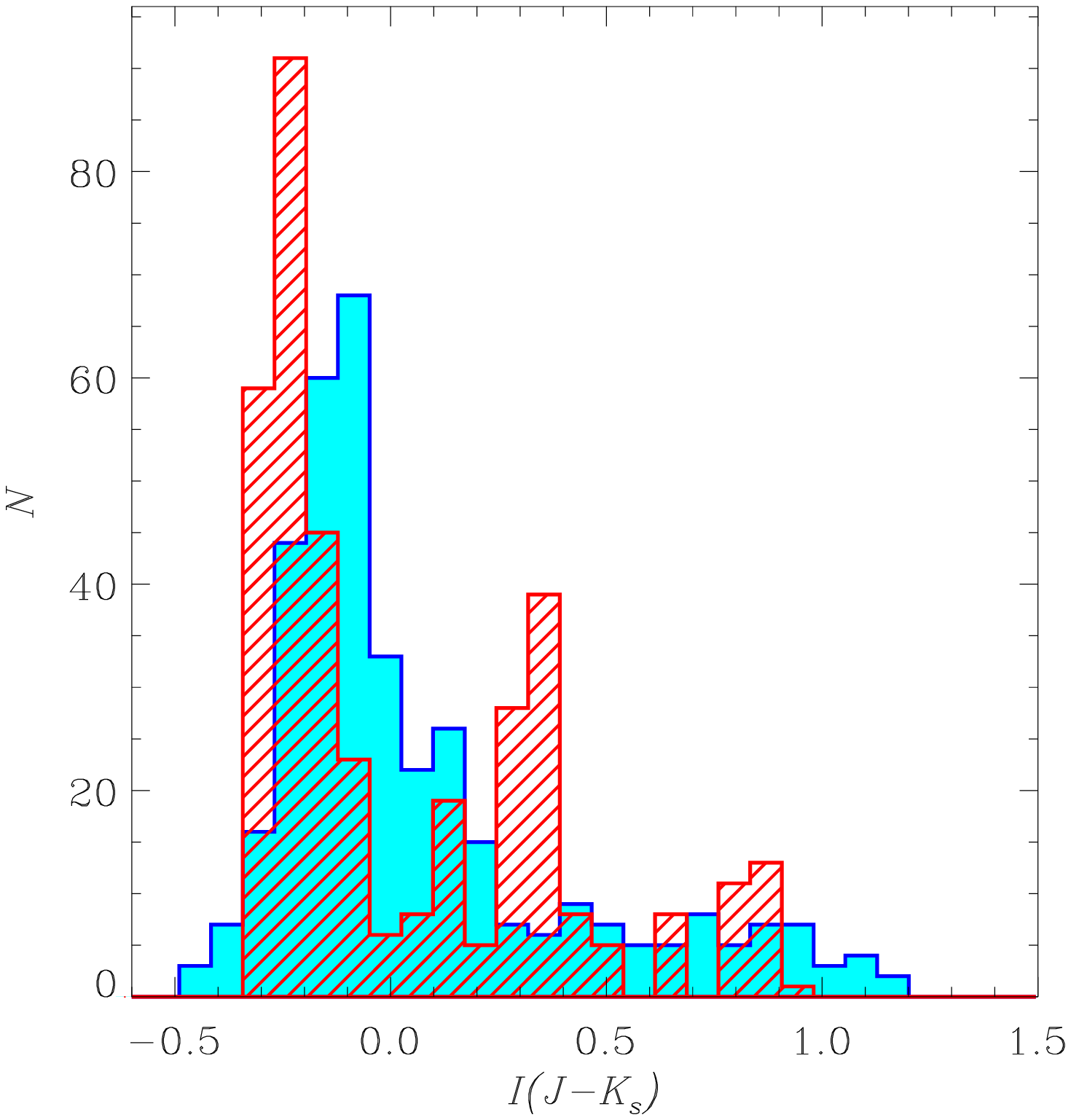}
}
\caption
{Observed colours of 650 Galactic open clusters compared to
theoretical colours computed from realistic and standard SSP models. The 
upper row is for $B$$-$$V$, the bottom one for $J$$-$$K_s$. The left column 
shows ``integrated colour $vs. \log t\,$'' diagrams, the right one the 
distributions of colour indices. Dots show Galactic open clusters from 
our sample. Black dots are young clusters used for constructing
the colour distributions, grey dots the older clusters. 
The curves are the model 
tracks constructed in this paper. The dashed curve is for the standard 
model, crosses show the realistic one. The observed colour distributions 
are shown as filled cyan histograms, the realistic model distribution 
as the hatched red ones.
}

\label{rev_fig}
\end{figure}

The observed clusters show a clear pattern, which at $\log t \lesssim 8.5$ is
represented by ``main sequence'' clusters (not containing red giants or supergiants).
The considerable redward spread of observed points cannot be attributed to
photometric errors; it is provided in this part of the diagram by a few
clusters containing yellow or red supergiants. For $\log t > 8.5$, one finds,
in general, clusters with red giants. The colour spread of older clusters is caused by
different evolutionary stages of red giants in clusters of different ages. The
depletion of the cluster sequence after $\log t\approx8.5$ can be explained
by either cluster mortality \citep[the typical lifetime of local clusters is
only around $\log t\approx8.5$, according to ][]{clupop} or selection
effects, which are more significant for older clusters.

In the right column of Fig.~\ref{sev_fig}, we compare the distributions of the observed cluster
colours with the GALEV predictions. We took the ages of the real clusters from 
\citet{intpar}, and calculated the integrated colours from the GALEV model.
The discrepancy between the observed colours and those from the standard model
is obvious. The local star clusters in our Galaxy are much bluer (by about 0.3 mag in the optical and about 0.8 mag in the near IR)
than ``theoretical'' clusters of the same age. We discuss this below in connection
with the results for the realistic model.

\section{The discreteness of the IMF and the SSP models}
\label{sec:descr}

As we have already mentioned, a possible explanation of the
observed disagreement is the continuity of the IMF adopted in standard
SSP models, which means that every small mass
interval of the given isochrone emits light according to its $T_{eff}$ and $\log g$.
The total flux that a ``theoretical'' cluster emits is a result of the 
integration over the isochrone, weighted by the given IMF.
It is immediately obvious that luminosity then scales with mass, once
the age (or the isochrone) is fixed. 
In the discrete approach, however, 
only points on the isochrone where stars of a respective mass
are actually present contribute. The random realisations of the IMF have statistical
fluctuations even for a smooth probability density distribution (i.e., smooth
IMF). At the high-mass end of the mass range, these fluctuations are particularly
large because of low number statistics, and produce gaps in the mass spectrum.
In the case of a single supergiant present in the
cluster at a given instant of time - a case frequently occurring in our local
Galactic clusters - the colour fluctuations might be especially
strong.
If the main sequence lifetime of the star closest in mass to the supergiant
exceeds the supergiant's lifetime, then after the supergiant's
death the less-massive star remains on the
main sequence, which makes the cluster considerably bluer than it was before.
The sparser the stellar population of a cluster is (i.e., the lower its total mass)
the stronger the colour fluctuations are in the course of its evolution.
After the most massive stars die, the red giant
population of a cluster reaches an equilibrium and the colour fluctuations of a
cluster begin to weaken.

Fig.~\ref{rev_fig} is the counterpart of Fig.~\ref{sev_fig} for a realistic
model of the scenario described above. The solid line with crosses and the dashed line
shown here correspond to the
discrete-IMF and to the continuous-IMF regimes, respectively, 
of the same model. The total initial
mass of the  model is assumed to be $M_c= 1000\, m_\odot$, and
$m_l=0.08\,m_\odot$. These parameters were chosen to provide a fit to the
colour distribution in the simplest case of our local young cluster population.
The ultimate upper mass limit $m_u=100\,m_\odot$ is defined
by the Padova grid of isochrones, and $m_u$ depends on both of the adopted $M_c$ and
$m_l$. For example, for $M_c=1000\,m_\odot$ and $m_l=0.01,\,0.1,\,$ and $1\,m_\odot$, the
upper mass limit to the simulated mass spectrum equals $m_u\approx
60,\,90,\, \mathrm{and}\, 100\,m_\odot$. In the latter case, $m_u$ is limited by
the adopted evolutionary grid.

For the standard models, this choice of parameters is however not important.
The derived evolutionary track depends on neither $M_c$ nor 
$m_l$. The upper mass limit affects only the early stages of the cluster evolution. The
only parameter influencing the standard model is the IMF slope $\alpha$, although
this influence is not strong. To change substantially the track,
one would need to increase the exponent $\alpha$ to 4.35. 
A flattening of the IMF with respect
to the Salpeter value of the slope does 
not have a strong effect on the colour-age
relation. We note that our continuous-IMF model agrees with the 
predictions of the
GALEV code and, hence, produces colours redder than the observed colours at $\log
t < 8.5$.

The realistic model with the above parameters produces a mass spectrum consisting
of 3532 stars. In the mass range
$5\,m_\odot>m>0.3\,m_\odot$, however, the number of stars is only 599. This is
of the same size as
the number of Pleiades members observed in this mass range, which
according to \citet{ple98} equals 780. Thus, the model appears to be
comparable to the Pleiades. The number of model stars with $m>5\,m_\odot$ is equal
to 20. So, before the model cluster reaches the age of the  Pleiades
($\log t\approx 8.1$), it has produced 20 red supergiants.
The twenty corresponding 
``RG-events'' are clearly seen in Fig.~\ref{rev_fig}. At all other times
(until $\log t\approx 8.1$), the cluster is seen as  a ``main-sequence cluster'' and
resides in the MS-domain of the diagram.
If we increase the lower bound $m_l$, the number of
massive stars increases and the fraction of ``RG-events'' slowly increases,
especially at older ages. But even at $m_l=1\,m_\odot$, the ``MS-clusters'' are
present in the track up to $\log t \approx 8.2$. On the other hand, decreasing $m_l$
diminishes the number of ``RG-events''. In principle, from the frequency of red
outliers in the ``colour versus $\log t$\,'' diagram, one can estimate 
indirectly the parameter $m_l$.

Although the above example provides only an explanation for the local cluster 
population of
typical mass of order of $10^3\, m_\odot$, it nevertheless illustrates the effect of
the discreteness 
on extragalactic clusters where the bulk of cluster data comes from
analysing the integrated light. 
For clusters of mass of the order of $10^5\,m_\odot$ \citet{dolkenn02} found that 
this effect is rather small.
This finding was interpreted by some workers \citep[e.g.,][]{bastea05} as a
justification for neglecting discreteness. Among other arguments,
\citet{bastea05} refer to the large masses of the observed extragalactic clusters, and
assume that the effect only produces a symmetric spread around the ``true'' sequence.
However, observations also find
extragalactic clusters with masses typical of Galactic clusters $\log
M_c=2-5$ \citep[see][]{moraea09}.
Since Fig.~\ref{rev_fig} shows that the
effect is highly asymmetric with respect to the standard SSP sequence,
a study of the effect over the entire range of cluster masses is appropriate.

\begin{figure}
   \centering
\resizebox{\hsize}{36mm}{
\includegraphics[clip]{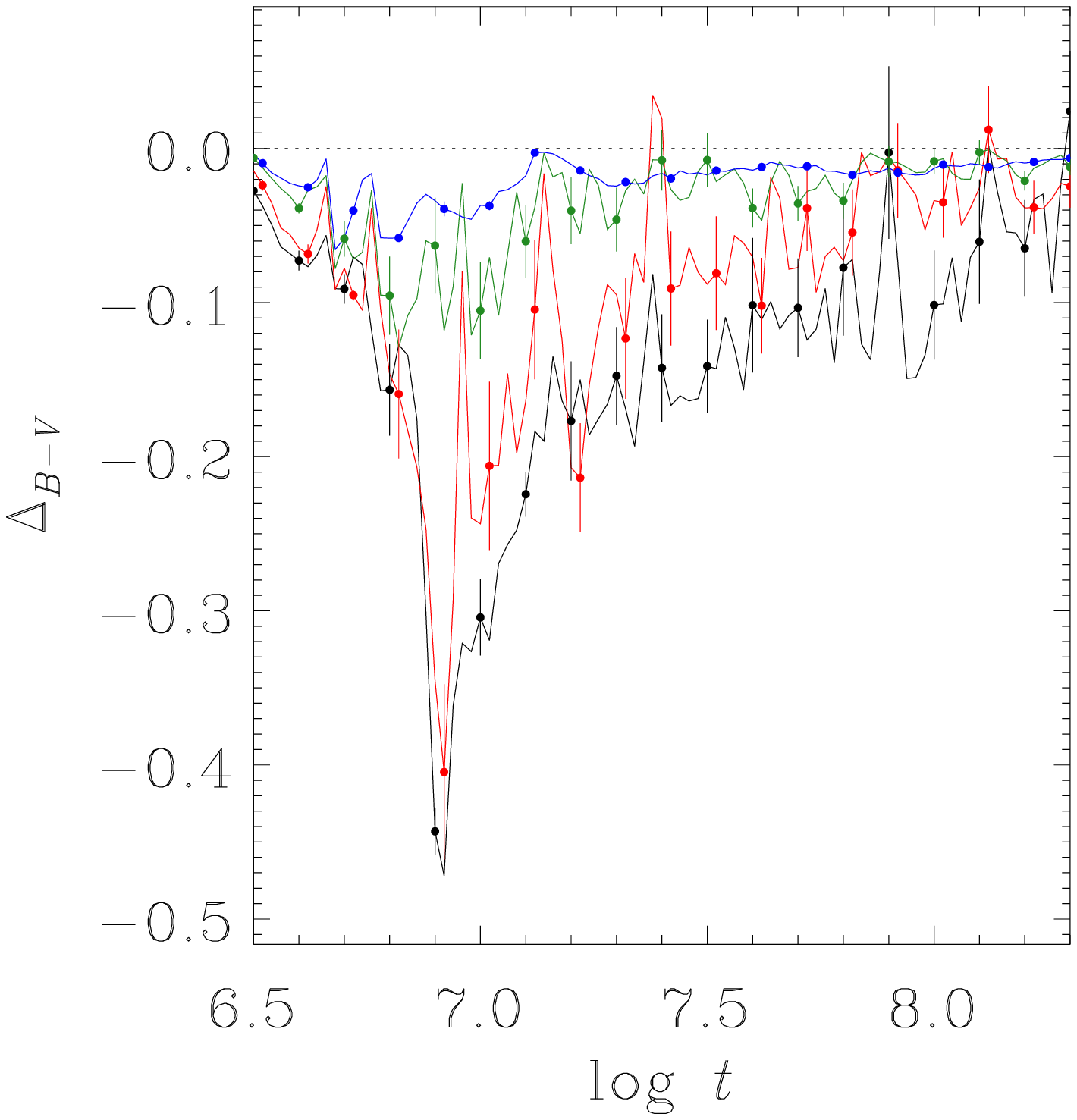}
\includegraphics[clip]{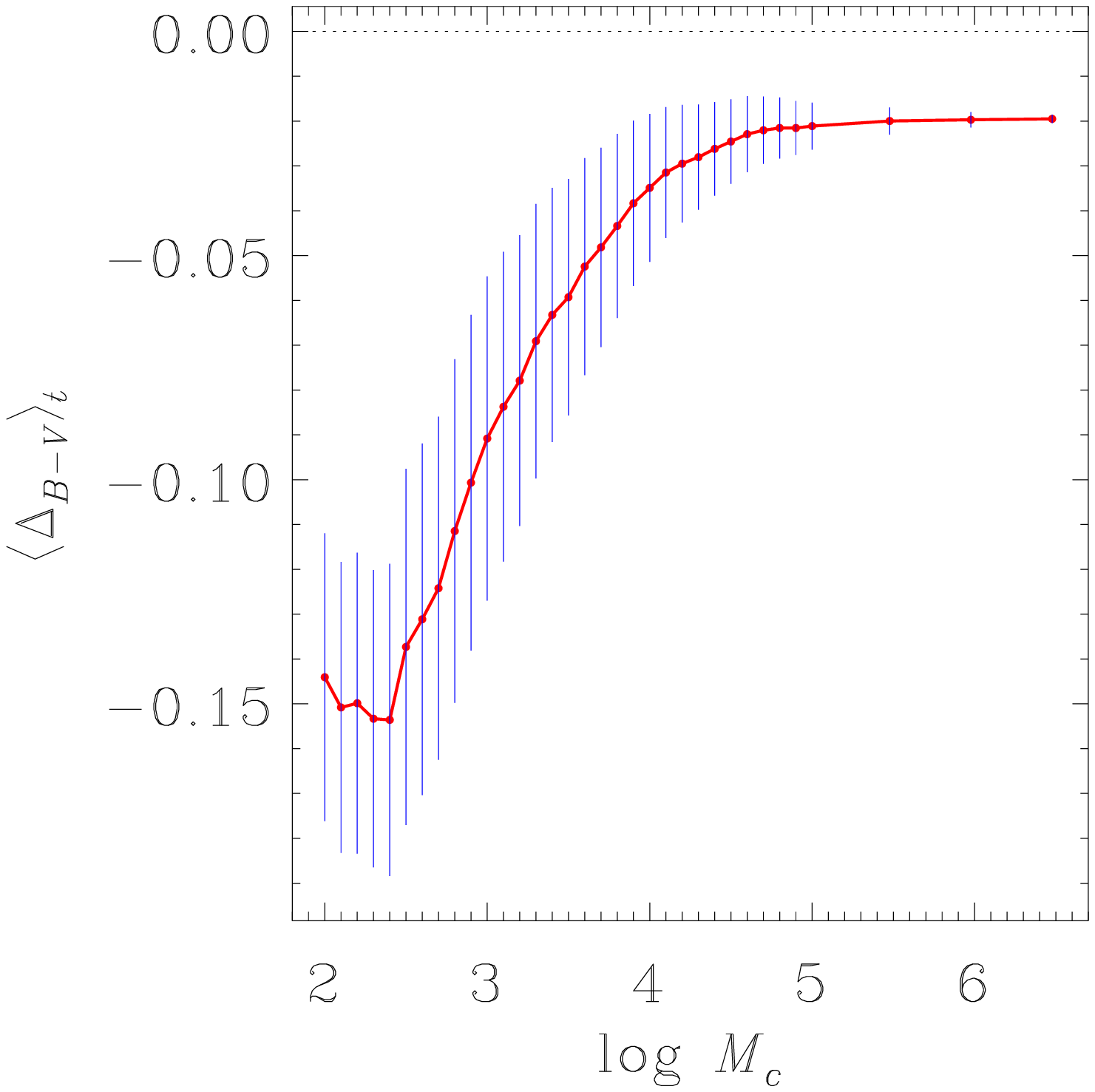}
}
\caption{The colour displacement versus age relations (left) for cluster models of
different masses. The curves correspond (from bottom to top) to $\log M_c=$ 2, 3, 4,
6.5. The vertical bars indicate the statistical uncertainty $\varepsilon$. In
the right panel, we show the displacement averaged over the age interval $\log t=
6.5-8.3$ as a function of cluster mass. The bars indicate the error in the
averaging.}
\label{rcol_fig}
\end{figure}

To estimate the dependence of the discreteness effect on cluster age and mass, we
have performed a series of Monte Carlo simulations. We considered a set of
cluster models with masses spread over the range $\log M_c=2.0-6.5$. For every
model of a particular mass at the starting point, the specific realisation of the
initial mass spectrum was sampled at random from the IMF. The evolution 
of the synthetic cluster was then followed up to  $\log t<8.3$.
At every time step,  we compared the colours of the
realistic model with its standard counterpart and computed the colour differences
$\delta=I(B$$-$$V)_{r}-I(B$$-$$V)_{s}$. Here the subscript $r$ 
represents ``realistic'', and
$s$ ``standard''. For every $M_c$, we compiled $N=50$ evolutionary
sequences of cluster models by considering different statistical realisations of the
initial mass spectrum, and evaluated at every time step the
corresponding average difference in colour
\[
\Delta = \sum\limits_{i=1}^N \delta_i/N,
\]
and its statistical uncertainty $\varepsilon$ computed to be 
\[
\varepsilon^2 = \sum\limits_{i=1}^N(\delta_i-\Delta)^2/(N-1)/N\,.
\]
In Fig.~\ref{rcol_fig}, we summarize the colour bias produced by neglecting the
IMF-discreteness effect, in the dependence on both the age and mass of a star cluster. 
The left panel shows the dependence of $\Delta$ on cluster age. The curves are for
different model cluster masses. The bars indicate the uncertainty
$\varepsilon$. The first conclusion from Fig.~\ref{rcol_fig} is 
that there is an overall negative bias.
As expected, the maximum effect is observed at $\log
t\approx 7$, when the first red stars appear in the upper, sparsely populated CMD
of a cluster. At younger ages, the effect diminishes since both realistic and
standard models are populated with main-sequence stars only, which have
approximately the same colour indices at the upper MS. At $\log t > 7$, the effect
slowly weakens, reflecting the CMD geometry (distance between the MS and
Hayashi limit). At $\log t \approx 8$, $|\Delta|$ becomes lower than 0.1
for all considered masses.
With increasing cluster mass, the effect in general diminishes. It is
interesting to note that even at $\log M_c=6.5$ the effect is small, but
significant, and at $\log t=6.7-7.0$ the colour shift is of the order of a few
hundredths of a magnitude. For masses $\log M_c > 4$,
$|\Delta|$  remains below 0.1 at all ages considered.

The explicit dependence of the discreteness effect on cluster mass, providing
the net-estimate of the effect for young clusters, can be derived if one
averages the $\Delta$ versus $\log t$ relations over the entire range of ages. The result
is shown in the right panel of Fig.~\ref{rcol_fig}. The effect
is considerable ($|\langle\Delta\rangle_t| \gtrsim 0.05$) for $\log M_c < 4$. At
higher masses it is small although significant.

For the other colours the behaviour is qualitatively similar, although the redder the
passband, the stronger the effect is. This is intuitively understandable. The
effect is caused by quantization, 
because red stars radiate proportionally more of their light at longer
wavelengths. For example, in the case of $I(J$$-$$K_s)$ the largest difference between
the realistic and the standard models is about $-0.9$ (cf. Fig.~\ref{rcol_fig}).
In a forthcoming paper, this issue will be discussed in more detail.

\section{Concluding remarks}
\label{sec:conc}

Observations of local Galactic open clusters have enabled us to measure their 
integrated magnitudes, masses,
ages, and reddening. Using this data set, we have found a remarkable
discrepancy between the observed colours and the predictions of SSP models.
The main reason for this disagreement is the neglect of the assumption of 
IMF-discreteness.
When this effect is taken into account, the model agrees adequately with 
the observations
and is even able to explain  the large colour spread observed
in the empirical colour-age relation in a natural way.

In which conditions is the effect of discreteness relevant? 
Since it is a consequence
of low-number statistics, it depends on the sparseness of the stellar
population in the upper CMD of a cluster where the bulk of light is emitted.
The density of the population depends primarily on cluster mass, and in
addition on cluster age, on [strange at first glance] the
lower mass limit of the stars formed, and to some degree on the slope of the IMF. 
In the context of this letter, other factors are not important.

According to our investigation, the \textit{systematic} offset between the continuous-
and discrete-IMF colours diminishes substantially but not completely at $\log t
>7.9$, at cluster masses $M_c> 10^5\,m_\odot$. At younger ages, it remains
present even in massive clusters, and for $M_c \leqslant 10^4$ it is larger than
0.1 mag in $(B$$-$$V)$. Only for very massive clusters ($M_c>10^6\,m_\odot$) 
and young ages ($\log t< 7.5$), the offset
falls below a typical
observational error. We note in passing that the effect is stronger
for redder passbands.
These findings are in good agreement with the theoretical 
forecast of \citet{cervino04}.
The immediate consequence of the application of a continuous-IMF approach 
to the SSP models
of stellar clusters is a systematic underestimate of reddening. Other problems
affect the age and mass determination based on evolutionary grids of these models
(at least for masses below $10^6\,m_\odot$).
Because of the
rather flat colour-age relation of young clusters, a systematic error of 
the order of 0.1 mag
produces an error in cluster age of about one order of magnitude.
On the other hand, there is a danger that the
existence of a large number of ``main-sequence'' (i.e., blue) clusters 
could be interpreted as evidence of a recent burst of star formation.
We can exclude such a burst having occured in our Galactic open clusters 
even if they are blue.
Finally, the
technique of the parameter determination should be changed to
incorporate the cluster flux variations caused by the discrete nature of the upper
IMF.

\begin{acknowledgements}
This study was supported by DFG grant 436 RUS 113/757/0-2, and RFBR grant 07-02-91566.
We thank Drs. Anders
and Ma{\'{\i}}z Apell{\'a}niz for drawing our
attention to the Cervi{\~n}o \& Luridiana paper.
\end{acknowledgements}

\bibliographystyle{aa}
\bibliography{clubib}

\end{document}